\documentstyle[version2,aps,preprint]{revtex}
\begin{document}

\begin{title}
Strong-Coupling Behavior of Two $t-J$ Chains\\
 with Interchain Single-Electron Hopping
\end{title}

\author{ Guang-Ming Zhang,\cite{gmz} Shiping Feng,\cite{sf} and Lu
Yu\cite{yu}}
\begin{instit}
International Center for Theoretical Physics, P.O.Box 586, 34100,
Trieste, Italy.
\end{instit}

\begin{abstract}
Using the fermion-spin transformation to implement spin-charge
separation of constrained electrons, a model of two $t-J$ chains with
interchain single-electron hopping is studied by abelian
bosonization. After spin-charge decoupling the charge dynamics  can be
trivially solved, while the spin dynamics is determined by a strong-coupling
fixed point where the correlation functions can be calculated explicitly.
This is a generalization of the Luther-Emery line for two-coupled
$t-J$ chains. The interchain single-electron hopping changes the asymptotic
behavior of the interchain spin-spin correlation functions and the
electron Green function, but their exponents are independent
of the coupling strength.
\end{abstract}


\newpage
An important issue of current interest is whether the peculiar properties
of one-dimensional (1D) Luttinger liquid (LL) \cite{haldane},\cite{and1}
will survive in two- and three-dimensions. The renormalization group (RG)
studies seemed to indicate  an instability of LL behavior with respect to
the interchain single-electron hopping (SEH) $t_{\perp}$ \cite{ws}. However,
Anderson suggested that {\it one should treat the intrachain correlations
exactly including the spin-charge separation before switching on}
SEH. Using the asymptotic Green functions
in 1D for a finite Hubbard $U$, he argued that SEH
is an irrelevant variable and named  this property as "confinement" of
the 1D Hubbard model \cite{and2}. His idea has stimulated several further
studies \cite{fpt}--\cite{sch}, most of which did not confirm his
conjecture in the strict sense. SEH is indeed renormalized to zero,
but e-e or e-h pair hopping is generated, which
drives the coupled chains towards a strong-coupling fixed point
corresponding to superconducting or density-wave states.
In fact, this type of instability was studied earlier in connection
with organic superconductors \cite{bg}. Nevertheless, this result is
not convincing because the validity of the perturbative RG
at strong-coupling fixed point  with large $U$ is questionable, as in the
single-impurity Kondo
problem. The Kondo physics is determined by the Wilson strong-coupling fixed
point
\cite{tou}. The poor-man's scaling \cite{ayh}
correctly directs the RG flow towards it, but
the calculation can not be justified by itself \cite{noz}.

In this paper, we consider two coupled $t-J$ chains, using a fermion-spin
transformation, proposed recently by Feng {\it et al}. \cite{feng}, where
the charge degrees of freedom  are described by spinless fermions, while
the spin degrees of freedom are represented by {\it hard-core} bosons, which
in  turn, can be expressed as  another type of spinless fermions via
Jordan-Wigner transformation.
The on-site local constraint for single occupancy is satisfied even in the
mean-field approximation (MFA) and the sum rule for physical electrons is
obeyed.  We combine this transformation with the abelian bosonization
technique \cite{haldane},\cite{lpef} to consider
the effect of SEH on the correlation
functions. After spin-charge decoupling the charge dynamics
can be solved trivially, while the spin dynamics can be mapped into
noninteracting spinless fermions. This
strong-coupling fixed point  is similar to the Luther-Emery line of the
single chain problem with back scattering \cite{le}.
We confirm that the spin-charge separation by itself does not
produce Anderson confinement \cite{fp}, \cite{fl}. Moreover, SEH changes
the asymptotic behavior of the
interchain spin-spin correlation functions and the electron Green
function, but their exponents are
independent of the coupling strength $t_{\perp}$.

We consider two coupled $t-J$ chains
\begin{eqnarray}
&&H=-t_{\parallel}\sum_{i,\sigma}(C_{1,i,\sigma}^{\dag}C_{1,i+1,\sigma}
+C_{2,i,\sigma}^{\dag}C_{2,i+1,\sigma}+h.c.)-
\mu\sum_{i,\sigma}(C_{1,i,\sigma}^{\dag}C_{1,i,\sigma}
+C_{2,i,\sigma}^{\dag}C_{2,i,\sigma}) \nonumber \\ && \hspace{1cm}
+2J\sum_i(\vec{S}_{1,i} \vec{S}_{1,i+1}+\vec{S}_{2,i} \vec{S}_{2,i+1})-
t_{\perp}\sum_{i,\sigma}(C_{1,i,\sigma}^{\dag}C_{2,i,\sigma}
+h.c.),
\end{eqnarray}
with local constraint $\sum_{\sigma}C_{i,\sigma}^{\dag}C_{i,\sigma}{\leq} 1$.
Here $C_{1,i,\sigma}^{\dag}(C_{2,i,\sigma}^{\dag})$   creates an
electron with spin $\sigma$ at site $i$ on chain 1 (2), and
$\vec{S}_{1,i}(\vec{S}_{2,i})$ is the corresponding electron spin operator;
$t_{\parallel}$ is the intrachain hopping and $\mu$ is the chemical
potential. The fermion-spin transformation of constrained electrons
\cite{feng}
$$C_{i,\uparrow}=P_{i}a_{i}S_{i}^{-}P_{i}^{\dag}, \hspace{1cm}
  C_{i,\downarrow}=P_{i}a_{i}S_{i}^{+}P_{i}^{\dag} $$
can implement the spin-charge separation without additional
constraints. Here $a_{i}$ and $a_{i}^{\dag}$ are "holon" (or
"electron" in the particle representation) operators,
represented by spinless fermions.
$S_{i}^{\pm}$ and $S_{i}^{z}$ are spinons
or pseudo-spin operators represented by $CP^1$ {\it hard-core} bosons,
different from the electron spin operators in Eq.(1). $P$ is a projection
operator removing the extra degrees of freedom in the $CP^1$
representation.
The anticommutation relations for constrained fermions $C_{i,\sigma}$
are strictly preserved. Moreover, the local constraint is
satisfied exactly. However, the projection
operator $P$ is cumbersome to handle and  in many cases, for example,
MFA, we can drop it, with very good results \cite{feng}.

To establish notations, consider first a single $t-J$ chain
\begin{eqnarray}\label{sc}
&& H_1=-t_{\parallel}\sum_{i}(a_{i}^{\dag} a_{i+1}+h.c.)(S_{i}^{+}
S_{i+1}^{-}+h.c.) -\mu\sum_i a_{i}^{\dag} a_{i} \nonumber \\ &&
 \hspace{1.5cm} +2J\sum_i (a_i^{\dag}a_i)
\vec{S_i} \vec{S}_{i+1}(a_{i+1}^{\dag} a_{i+1}).
\end{eqnarray}
 As shown in \cite{feng}, using the above fermion-spin representation, the
Jordan-Wigner transformation
$S_{i}^{+}=f_{i}^{\dag}e^{i\pi\sum_{l<i}{f_{l}^{\dag}f_{l}}},
S_{i}^{-}=(S_{i}^{+})^{+},
 S_{i}^{z}=f_{i}^{\dag}f_{i}-{\frac{1}{2}}$,
and MFA, one finds the ground state energy and
gapless spinon and holon
spectra, in good agreement with the exact solution \cite{shiba}. However, to
obtain correct exponents for correlation functions, one has to go beyond
the MFA, taking into account holon-spinon interactions.
Following Weng {\it et al}. \cite{weng}, this can be done by
"squeezing out" holes from the spin chain, {\it i.e.}, to replace
 $ a_{i}^{\dag}a_{i+1}(f_{i}^{\dag} f_{i+1} +f_{i} f_{i+1}^{\dag})$
by $ a_i^{\dag}a_{i+1}$ wherever there is a hole at site $i$ and introducing
the "string operators" which in our case are given by \cite{feng}
$$ C_{i,\uparrow}=[a_{i}e^{i{\pi}(N-\sum_{l>i}a_{l}^{\dag}a_{l})}]
[f_{i}e^{{-}i{\pi}\sum_{l<i}a_{l}^{\dag}a_{l}}],$$
$$C_{i,\downarrow}=
[a_{i}e^{i{\pi}(N+\sum_{l>i}a_{l}^{\dag}a_{l})}]
[f_{i}^{\dag}e^{i{\pi}\sum_{l<i}a_{l}^{\dag}a_{l}}].$$
\noindent
In the resulting Hamiltonian, the "holon" part is free and can be easily
bosonized, while the spinon part
is an antiferromagnetic Heisenberg spin-1/2 chain, which can
also be bosonized and reduced
to a  standard 1+1 quantum sine-Gordon (SG) model \cite{nijs}
\begin{equation}\label{ss}
H_{1,s}={\int}dx{\left(\frac{v_{s}K_{s}}{2}{\Pi}^{2}+\frac{v_{s}}{2K_{s}}
\left(\bigtriangledown\varphi\right)^{2}-{\frac{2v_{s}K_{s}^{2}k_{F}^{2}}
{(2\pi\alpha)^2}}cos{\sqrt{16\pi}\varphi}\right)},
\end{equation}
where $\alpha$ is an ultroviolate cut-off, while the boson field $\varphi$
describes the low-energy excitations of  spinons, $\Pi$ is its conjugate
momentum with a commutation relation $[\varphi(x),\Pi(x')]=i\delta(x-x')$.
The spinon velocity is
$v_{s}=2J\left[(1-{\delta})^{2}-\left(\frac{sin{\delta\pi}}{\pi}\right)^2
\right]{\sqrt{1+\frac{4}{\pi}}}$, with $\delta$ as doping concentration.
The
parameter determining the exponent of the spinon correlation function is
$K_{s}=(1+\frac{4}{\pi})^{-1/2}$, which should be independent of $\delta$,
and our result for $K_{s}$  is slightly away from the
exact value derived for half-filling \cite{nijs}, \cite{yyb}.
In principle, the
abelian bosonization is exact only at $J_{z}/J_{\perp}{\approx}0$
for the Heisenberg spin-1/2 chain \cite{lpef}. However, the exact
Bethe-ansatz solution  does not show any singularities for
 $-1<J_{z}/J_{\perp}<1$, and the isotropic antiferromagnetic coupling
$J_{z}/J_{\perp}=1$ is described by this fixed point \cite{yyb}. On
the other hand, the 1+1 SG model with $A$cos$\beta\varphi$ has only
one weak-coupling fixed point for small $A$ at $\beta^{2}=8\pi$
\cite{coleman}. Thus we can associate the cosine interaction (\ref{ss}) with
this fixed point of the SG model in order to rectify $K_{s}$
to be $1/2$ after rescaling $ \Pi\rightarrow\sqrt{K_s}\Pi$, $
\varphi\rightarrow\frac{\varphi}{\sqrt{K_s}}$. Since the fixed point of SG
Hamiltonian under RG for $\beta\geq 8\pi$ corresponds to the vanishing of
the cosine term \cite{coleman}, we can easily calculate the asymptotic
behavior of
the spin-spin correlation functions and the electron Green
functions, in good agreement with exact results \cite{shiba}, {\it e.g.,}
\begin{equation}\label{ssc}
S(x_i-x_j,t)\sim \frac{cos{2k_F(x_i-x_j)}}
{\left[(x_i-x_j)^2-(v_ht)^2\right]^{\frac{1}{4}}
\left[(x_i-x_j)^2-(v_st)^2\right]^{\frac{1}{2}}},
\end{equation}
\begin{eqnarray}\label{cec}
\langle TC_{i,\sigma}(t)C_{j,\sigma}^{\dag}(0)\rangle \nonumber \\ &&
\noindent \sim \frac{e^{ik_F(x_i-x_j)}}
{\left[(x_i-x_j)^2-(v_ht)^2\right]^{\frac{1}{16}}
\left[(x_i-x_j)-(v_st)\right]^{\frac{1}{2}}
\left[(x_i-x_j)-(v_ht)\right]^{\frac{1}{2}}},
\end{eqnarray}
where the holon velocity is the exact value
$v_{h}=2t_{\parallel}sin{\delta\pi}$, and
$k_F=\frac{\pi}{2}(1-\delta)$.

Now consider two coupled chains and use the MFA to decouple the
interchain holon-spinon interaction.
The Hamiltonian (1) is reduced to the following form
$H=H_{h}+H_{s}$, and
\begin{eqnarray}
&& H_{h}=-t_{\parallel}\sum_{i}(a_{1,i}^{\dag}a_{1,i+1}+a_{2,i}^{\dag}
a_{2,i+1}+h.c.)-t_{\perp}{\eta}_{1}\sum_{i}(a_{1,i}^{\dag}a_{2,i}+h.c.)
\nonumber\\ && \hspace{1.5cm}
 -\mu\sum_i(a_{1,i}^{\dag} a_{1,i}+a_{2,i}^{\dag} a_{2,i}),
\end{eqnarray}
\begin{equation}
 H_{s}=2J^{eff}\sum_{i}(\vec{S}_{1,i}\vec{S}_{1,i+1}+\vec{S}_{2,i}
\vec{S}_{2,i+1})-t_{\perp}{\eta}_{2}\sum_{i}(S_{1,i}^{+}S_{2,i}^{-}+
h.c.),
\end{equation}
where we have defined two MF order parameters
${\eta}_1$ and $ {\eta}_{2}$.

The holon Hamiltonian is trivially diagonalized by introducing
 $A_{k}={\frac{1}{\sqrt{2}}}(a_{1,k}+a_{2,k})$ and
$B_{k}=-{\frac{1}{\sqrt{2}}}(a_{1,k}-a_{2,k})$
with excitation energies
$ \varepsilon_{k}^{A}=-2t_{\parallel}cos{k}-t_{\perp}$ and
$ \varepsilon_{k}^{B}=-2t_{\parallel}cos{k}+t_{\perp}$, respectively,
where we assume $\eta_1\approx 1$, as will be confirmed later.
The SEH splits the original holon excitation
spectrum by $2t_{\perp}$ and in the low doping case for a finite
value
$t_{\perp}>t_{\parallel}(1-cos{2\delta\pi})$,
only the upper band has vacancies and the lower band is fully
occupied. The above condition on $t_{\perp}$ is usually satisfied.
Thus, it is
easy to find the self-consistent value ${\eta}_2=-(1-\delta)$, as well as
the interchain
holon correlation functions using the abelian bosonization technique
\cite{haldane}
\begin{equation}
\langle e^{{\mp}i{\pi}\sum_{l<i}a_{1,l}^{\dag}(t)a_{1,l}(t)}
e^{{\pm}i{\pi}\sum_{l<j}a_{2,l}^{\dag}(0)a_{2,l}(0)}\rangle
{\sim}
{\left[(x_{i}-x_{j}^{'})^{2}-(v_{h}t)^{2}\right]^{-\frac{1}{16}}},
\end{equation}
\begin{eqnarray}
&&
\langle e^{{-}i{\frac{\pi}{2}}\sum_{l<i}a_{1,l}^{\dag}(t)a_{1,l}(t)}
a_{1,i}(t)a_{2,j}^{\dag}(0)
e^{i{\frac{\pi}{2}}\sum_{l<j}a_{2,l}^{\dag}(0)a_{2,l}(0)}\rangle
\nonumber\\ && {\sim}
{\frac{e^{ik_{F}^{B}(x_{i}-x_{j}^{'})}}{
{\left[(x_{i}-x_{j}^{'})^{2}-(v_{h}t)^{2}\right]^{\frac{1}{64}}}
{\left[(x_{i}-x_{j}^{'})-(v_{h}t)\right]^{\frac{1}{2}}}}},
\end{eqnarray}
where $k_F^B=k_F-\frac{t_\perp}{v_h}$.
Due to the single occupancy constraint, the situation here is simpler than
the weak-coupling case where both bands have to be
considered \cite{fpt}-\cite{sch}.

 The spinon part can be reduced to the following form:
\begin{eqnarray}\label{tcs}
&& H_s = \int dx  [\frac{ v_s K_s} { 2} \Pi_1^2 +
\frac{ v_s} {2K_s} (\bigtriangledown \varphi_1)^2
+\frac{v_s K_s} { 2} \Pi_2 ^2 +\frac{v_s}{ 2K_s}
(\bigtriangledown \varphi_2 ) ^2
\nonumber\\&& \hspace{2cm} + \frac{ (1-\delta ) t_\perp} { (\pi\alpha )^2}
\cos ( \sqrt{\pi}
(\tilde{\varphi}_1 -\tilde{\varphi}_2 ) )],
\end{eqnarray}
where ${\tilde{\varphi}}(x)$ is the dual
field of ${\varphi}(x)$, and is defined by
$\frac{\partial{\tilde{\varphi}}}{\partial{x}}=\Pi$ and
$ \tilde{\Pi}=-\frac{\partial{\varphi}}{\partial{x}}$. Note the
difference of (\ref{tcs})
 from  (\ref{ss}), where no dual fields are involved. Introducing
symmetric and antisymmetric combinations of spinon fields:
${\varphi}_{S}={\frac{1}{\sqrt{2}}}({\varphi}_{1}+{\varphi}_{2})$,
${\varphi}_{A}={\frac{1}{\sqrt{2}}}({\varphi}_{1}-{\varphi}_{2})$,
(\ref{tcs}) can be rewritten as:   $H_{s}=H_{s}^{S}+H_{s}^{A}$,
where
\begin{equation}
H_{s}^{S}={\int}dx{\left(\frac{v_{s}K_{s}}{2}{\Pi}_{S}^{2}+
\frac{v_{s}}{2K_{s}}\left(\bigtriangledown{\varphi}_{S}\right)^{2}\right)},
\end{equation}
\begin{equation}
H_{s}^{A}={\int}dx{\left({\frac{v_{s}K_{s}}{2}}{\Pi}_{A}^{2}+
{\frac{v_{s}}{2K_{s}}}{\left(\bigtriangledown{\varphi}_{A}\right)}^{2}+
{\frac{(1-\delta)t_{\perp}}{(\pi\alpha)^{2}}}{cos(\sqrt{2\pi}
{\tilde{\varphi}_{A}})}\right)}.
\end{equation}
The symmetric part $H_{s}^{S}$ is a LL
 with the same parameters $v_s$ and $K_s$ as for a single
chain. As for the antisymmetric part, a rescaling and use of
self-duality of the non-interacting part lead to the following 1+1
 quantum SG model
\begin{equation}
H_s^A=\int dx \left(\frac{v_{s}}{2}[\tilde{\Pi}_{A}^{2}+
\left(\bigtriangledown\tilde{\varphi}_{A}\right)^{2}]+
\frac{(1-\delta)t_{\perp}}{(\pi\alpha)^{2}}cos(\sqrt{\frac{2\pi}{K_{s}}}
\tilde{\varphi}_{A})\right).
\end{equation}
With the corrected $K_{s}=\frac{1}{2}$, the coupling strength of the SG
model is ${\beta}^{2}=4\pi$. This is nothing but a free
massive Thirring model \cite{coleman} with a mass gap in the excitation
spectrum ${\bigtriangleup}_s\approx 2(1-\delta)t_{\perp}$. It is known that
$t_{\perp}$ is a relevant variable
in the range  $0<{\beta}^2<8\pi$. Some years ago, Haldane \cite{h}
considered the renormalization of the
Bethe ansatz equations for the massive Thirring model, equivalent to
the 1+1 SG model \cite{bt}. He  extracted a quantum
fluctuation parameter that controlled the correlation functions of this
model, and found that at ${\beta}^2=4\pi$ the renormalization of the model
stops and it corresponds to a free spinless fermion field.
This means that ${\beta}^2=4{\pi}$ is just the strong-coupling fixed point ,
analogous to
the Toulouse limit of the single-impurity Kondo
problem \cite{tw}. It is the fixed point that controls the properties of
this model in the whole region $0<{\beta}^2<8{\pi}$.
It is remarkable that after
correcting the $K_s$ value using the Bethe ansatz solution for the single
chain, we end up {\it exactly at this fixed point for two coupled t-J chains}.
If we did not rectify
the parameter $K_s$ in the absence of the SEH, the coupling
strength of the above SG model would be at
$({\beta}^{'})^2<4\pi$, and it should be
renormalized to strong-coupling fixed point  ${\beta}^2=4\pi$,
while the parameter $K_s$ is renormalized to
${\frac{1}{2}}$. In the end, the spinon correlation functions of the single
$t-J$ chain could still have correct asymptotic behavior.

In the weak coupling approach, it was also found that the ground state of
a single chain is unstable with respect to SEH and the
two-coupled chains are driven to a strong-coupling fixed point  corresponding
to ${\beta}^{2}=4\pi$ with opening a gap in one of the spinon excitation
spectra \cite{fl}, which is equivalent to the Luther-Emery line for a single
chain with back scattering. However, for a finite $t_{\perp}$, the
spin
and charge degrees of freedom are still coupled, and the renormalizaion
process cannot be carried out in the perturbative approach.
Therefore, it is  not
possible to calculate the correlation functions at this strong-coupling fixed
point .
On the contrary, within our
strong-coupling approach, we end up exactly at this strong-coupling fixed point
 and the spinon
excitations reduce to two
modes, one is the original LL branch for the single chain, while the other
is a massive free fermion branch, corresponding to the soliton gas of the
quantum SG model.
After convolution, the interchain spin-spin correlation function and
electron Green function for two-coupled chains can be
calculated as
\begin{equation}
S(x_{i}-{x_{j}^{'}},t){\sim}{\frac{cos{2k_{F}}(x_{i}-{x_{j}}^{'})}
{\left[(x_{i}-{x_{j}}^{'})^{2}-(v_{h}t)^2\right]^{\frac{1}{16}}
\left[(x_{i}-{x_{j}}^{'})^{2}-(v_{s}t)^2\right]^{\frac{1}{4}}}},
\end{equation}
\begin{eqnarray}
&&
\langle TC_{1,i,\sigma}(t)C_{2,j,{\sigma}}^{\dag}(0)\rangle
\nonumber\\&& {\sim}{\frac{e^{ik_{F}^{B}(x_{i}-{x_{j}}^{'})}}
{\left[(x_{i}-{x_{j}^{'}})^{2}-(v_{h}t)^2\right]^{\frac{1}{64}}
\left[(x_{i}-{x_{j}^{'}})-(v_{s}t)\right]^{\frac{1}{4}}
\left[(x_{i}-{x_{j}}^{'})-(v_{h}t)\right]^{\frac{1}{2}}}}.
\end{eqnarray}
The parameter $\eta_1\approx 1$, as mentioned earlier. As compared with
 (\ref{ssc}) and (\ref{cec}) for a single $t-J$ chain, the SEH has
generated new exponents,
{\it independent of} $t_\perp$. The singularity is weaker due to the
presence of a gap in one of the excitation branches.
The "spinon"  exponent is $-\frac{1}{2}$ instead of $-1$ for the spin-spin
correlation function, while it is $-\frac{1}{4}$ instead of
$-\frac{1}{2}$ for the electron Green function. The singularity
due to holons is also weakened because of the hybridization of two
chains and the single occupancy constraint.
The gap in the excitation spectrum for the antisymmetric spinon
field $\tilde{\phi}$ leads to an exponential decay for
$\langle\tilde{\phi}(x,t)\tilde{\phi}^{\dag}(0,0)\rangle$, but a constant
contribution to the correlation functions for the fermionic fields.

To summarize, we have found strong-coupling fixed point  controlling
the behavior of two coupled $t-J$ chains and have calculated explicitly
the interchain spin-spin correlation function and the electron
Green function. Our work has reconfirmed some results of the previous
weak-coupling studies, namely,
(i) the spin-charge separation does not produce by itself Anderson
confinement;
(ii) the exponents of the interchain correlation functions are changed
due to the presence SEH
\cite{fp},\cite{fl}.
However, there are significant differences between these two approaches.
(i) The weak-coupling approach indicates the existence of a strong-coupling
fixed point , but can not
provide a valid calculation scheme at that fixed point.
(ii) The spin-charge separation for a single chain in the weak-coupling
sense in general does not guarantee the spin-charge separation for
two coupled chains (except for a special $g_4$-ology model \cite{fp}), so
it is not possible to calculate explicitly the
correlation functions. The situation here is similar to the single chain
problem. The correlation exponents depend on the interaction strength in
the weak-coupling limit, while it is independent of interaction strength in
the strong-coupling limit; the spin-charge separation is valid in
the sense of "almost complete factorization" of the wave function in the
large-U limit \cite{shiba}. Here we use the spin-charge separation in the
same sense and it can be thus justified.

{\it Acknowledgements}. The authors are very grateful to Dr. X. Y. Zhang for
several enlightening comments. They appreciate helpful discussions with
A. Parola and E. Tosatti and thank A. Finkel'stein for sending Ref.9
prior to publication. They also acknowledge the support of ICTP for
these studies.

\end{document}